# Pulmonary embolism identification in computerized tomography pulmonary angiography scans with deep learning technologies in COVID-19 patients

Chairi Kiourt[1], Georgios Feretzakis[2,3], Konstantinos Dalamarinis[4], Dimitris Kalles[2], Georgios Pantos[4], Ioannis Papadopoulos[4], Spyros Kouris[4], George Ioannakis[1], Evangelos Loupelis[5], Petros Antonopoulos[4], Aikaterini Sakagianni[6]

[1]Athena-Research and Innovation Center in Information, Communication and Knowledge Technologies, Xanthi, Greece.
{chairiq, gioannak}@athenarc.gr

[2]School of Science and Technology, Hellenic Open University, 26335 Patras, Greece.
georgios.feretzakis@ac.eap.gr, kalles@eap.gr

[3]Department of Quality Control, Research and Continuing Education, Sismanogleio General Hospital, 15126 Marousi, Greece.
quality@sismanoglio.gr

[4]Department of Radiology, Sismanogleio General Hospital, 15126 Marousi, Greece.
kodalmar@gmail.com, gpantos1@gmail.com, ioannispapadopoulos@yahoo.com, spgecouris7338@gmail.com, petrosaantonopoulos@gmail.com

[5]IT Department, Sismanogleio General Hospital, 15126 Marousi, Greece.
v_loupelis@sismanoglio.gr

[6]Intensive Care Unit, Sismanogleio General Hospital, 15126 Marousi, Greece.
sakagianni@sismanoglio.gr

**Abstract**

The main objective of this work is to utilize state-of-the-art deep learning approaches for the identification of pulmonary embolism in CTPA-Scans for COVID-19 patients, provide an initial assessment of their performance and, ultimately, provide a fast-track prototype solution (system). We adopted and assessed some of the most popular convolutional neural network architectures through transfer learning approaches, to strive to combine good model accuracy with fast training. Additionally, we exploited one of the most popular one-stage object detection models for the localization (through object detection) of the pulmonary embolism regions-of-interests. The models of both approaches are trained on an original CTPA-Scan dataset, where we annotated of 673 CTPA-Scan images with 1,465 bounding boxes in total, highlighting pulmonary embolism regions-of-interests. We provide a brief assessment of some state-of-the-art image classification models by achieving validation accuracies of 91% in pulmonary embolism classification.

Additionally, we achieved a precision of about 68% on average in the object detection model for the pulmonary embolism localization under 50% IoU threshold. For both approaches, we provide the entire training pipelines for future studies (step by step processes through source code). In this study, we present some of the most accurate and fast deep learning models for pulmonary embolism identification in CTPA-Scans images, through classification and localization (object detection) approaches for patients infected by COVID-19. We provide a fast-track solution (system) for the research community of the area, which combines both classification and object detection models for improving the precision of identifying pulmonary embolisms.

**Keywords**: Pulmonary Embolism, COVID-19, Deep Learning, CTPA-Scans, Image classification, Object Detection,

## 1. Introduction

The thromboembolic disease occurs as a common complication in patients with severe COVID-19 disease leading to increased morbidity and mortality [1, 2]. Many meta-analyses have reported a Pulmonary Embolism (PE) rate of 16.5% in patients with severe COVID-19 infection, reaching 24.7% in more critically ill patients hospitalized in the ICU [3, 4]. Other studies report a higher rate of up to 37.1% [5]. In addition, patients with COVID-19 disease have a higher incidence of in situ thrombosis or microvascular thrombosis [6, 7], which is why they are mainly found in peripheral vessels (subsegmental branches). The clinical diagnosis of PE is extremely difficult because there are neither specific clinical signs/symptoms nor pathognomonic laboratory tests, therefore the combination of Computed Tomography Pulmonary Angiography (CTPA) with experts' (e.g. Radiologists) inspection of the area remains the gold standard for diagnosis [8]. CTPA-Scans consists of hundreds of images, where each image represents one slice of the lung, and the identification of PE with high clinical accuracy is time-consuming and complicated, mainly due to a high number of false-positive results [9]. Studies have shown that there may be a discrepancy of up to 13% in the CTPA assessment of PE between experienced and less experienced radiologists in chest imaging [10, 11]. Additionally, there is a lot of pressure on hospitals to provide 24/7 services for fast CTPA examinations and timely and accurate notification of results to the referring doctor [12, 13].

The applications of Deep Learning (DL) are promising in medical imaging on plain chest X-rays, Computed Tomography (CT) and Magnetic Resonance Imaging (MRI) [14] in various published studies [15], including COVID-19 diagnosis [16]. DL to automatically highlight PE in CTPA remains a major challenge compared to other radiological applications and presents many difficulties for various reasons [17]. The use of automated DL approaches to diagnose PE on CTPA could have excellent clinical application mainly because the definitive diagnosis of PE is made only by CT imaging and no further testing is required [18, 19]. The challenge is even greater in patients with COVID-19 infection due to the coexistence of multiple lung lesions that produce false-positive results.

Figure 1 depicts three different samples of CTPA-Scan images (from COVID-19 patients), by highlighting with red arrows the presence of important Regions-of-Interest (RoI) with PE. It is clearly a very challenging task for Computer Vision (CV) and Machine Learning (ML) scientists, because the images are low resolution, low quality and present high complexity information risen from a single channel (grayscale).

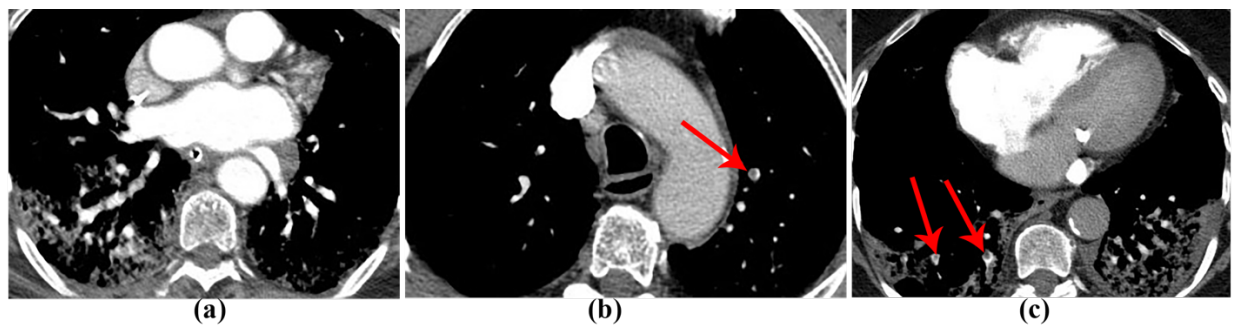

Figure 1. CTPA-Scan image sample from patients infected by COVID-19: a) without PE, b) with PE, an easy-case for experts and c) with PE, a difficult-case for experts.

By searching the literature under the umbrella terms of DL, CV, CTPA-Scans and PE we present some recent state-of-the-art related works, highlighting the importance and challenges of this research domain.

Rucco et al. [20] introduced the Neural Hypernetwork, an integrative approach based on Q-analysis with ML. The main objective of the approach is to improve PE diagnosis, while reducing the number of CTPA-Scans required for confirming the diagnosis. The experimental phase involved data from 28 diagnostic features of 1427 people considered to be at risk of PE and reached a satisfactory recognition rate of 94%. Huang et al. proposed PENet [18], a scalable deep-learning model for automated diagnosis of PE using volumetric CT imaging. The model consists of 77 3D convolutional neural layers and is pretrained on the Kinetics-600 dataset and fine-tuned on a retrospective CTPA-Scans dataset collected from an academic institution. PENet's performance was evaluated on data provided from two distinct institutions: one as a hold-out dataset from the same institution and one collected from another institution to evaluate the model's generalizability to an unrelated population dataset. It achieved an AUROC (Area Under Receiver operating characteristic Curve) of 0.84 with a false positive rate standard deviation of 2±0.02 on detecting PE on the hold out internal test-set and 0.85 standard deviation of 2±0.03 on external data. Rajan et al. [21] propose a two-stage detection pipeline designed exclusively using 2D CNNs, wherein the candidate generation state utilizes a novel context-augmented U-Net and the classifier stage employs a simple 2D Conv-LSTM model coupled with multiple instance learning. Within the first stage, CT volumes are processed to produce a mask that identifies candidate regions that are likely to correspond to embolism regions, while in the latter one, those masked regions are utilized to perform the actual detection. The approach achieves AUC scores of 0.94 on the validation set and 0.85 on the test-set, while the proposed approach has produced, even with a substantially smaller number of parameters and with no pre-training state-of-the-art detection, positive results on a challenging and large-scale real-world dataset. Tajbakhsh et al. [22] investigated the possibility of a unique PE representation, coupled with CNNs, that led to an accuracy increase of PE CAD system for PE CT classification.

Through studying the aforementioned related works and many other publications, it is highlighted that there is a lack of systems/algorithms that not only classify an image of a patient with PM, but also localize the suspect positions/RoI, under the noise (image quality affect) of COVID-19. This issue can be considered as the main motivation of this work with huge scientific and social impact. Differentiating from previous related works, the main contributions of our work are:

- The development of a dataset composed of multiple CTPA-Scan images of COVID-19 patients for classification experiments
- The annotation of RoIs in the CTPA-Scan images, with bounding boxes for the determination of the locations of the suspect RoI of PE, for Object Detection (OD) experiments.
- A short assessment of some state-of-the-art deep learning classification model, through Transfer Learning (TL) approaches, for the determination of the most accurate one in classification of CTPA-Scan images presenting PE.
- The utilization of one of the most fast and accurate OD model for the localization of the RoI of PE in CTPA-Scan images.
- Very accurate DL models for the identification of PE, achieving about 91% classification validation accuracy and about 68% Average Precision (AP) for the OD.
- The development of a fast-track prototype solution (system) for the research community of the area, which combines both classification and OD models for more precise identification of the PE.

## 2. Materials and methods

In this section, we present and analyze the new images/dataset risen from CTPA-Scans and also the two adopted DL approaches (two stages of experiments) for classification and localization of the PE in CTPA-Scan images. The source code for the training of DL models is provided through a source code management repository[1].

### 2.1 Dataset description

Nowadays, there are not many CTPA-Scan datasets presenting PE [9] and almost none of these datasets come from COVID-19 patients and neither are annotated with bounding boxes, highlighting the RoIs of PE, for localization/OD studies. For this reason, we created an original dataset and annotated the images based on the requirements of state-of-the-art OD models.

---

[1] https://tinyurl.com/pulemb-ctscans-code

This research was conducted in a public tertiary care hospital in Greece and it has been approved by the Institutional Review Board of Sismanogleio General Hospital. All patients' CTPA images used in this research study were anonymized and reviewed by two experienced radiologists (with 14 years and 23 years respective experience in thoracic imaging) for the presence of PE, from the Picture Archiving and Communication System (PACS) of the Radiology Department. All COVID-19 patients were diagnosed based on RT-PCR testing. The dataset included 19 patients diagnosed with COVID-19 with PE presence, 3 non-COVID-19 patients with PE presence, and 10 patients with COVID-19 with the absence of PE. For the classification of the PE, we used 673 images on each class (PE presence, "Yes" class / absence of PE, "No" class) total 1346 images. The dataset for the localization (object detection) training of the PE detection included 573 CTPA-Scan images from 22 patients with 1239 bounding boxes ranging from 1 to 8 bounding boxes/image with an average of 2.16 bounding boxes/image. Additionally, the test dataset for the localization process comprises of 100 CTPA images with 226 bounding boxes ranging from 1 to 7 bounding boxes/image and with an average 2.26 bounding boxes/image. In total, the dataset for PE detection included 673 images with 1465 bounding boxes with an average of 30.59 images/patient. The images from a CTPA-Scan of a patient are selected by expert radiologists, who checked to ensure that between images of the same patient there was an estimated difference of at least about 20% in visible morphology of the PE.

At this point, it should be highlighted that the creation of such dataset is a very difficult and time-consuming process and especially during the COVID-19 pandemic, where all medical experts who work in hospitals are under incredible pressure.

## 2.2 Image-based identification of pulmonary embolism

The scope of this experiment is to identify images from a CTPA-Scan of a patient with the existence of PE, through a classification approach. To get a more accurate model, with faster training, we adopted TL methods through fine-tuning approaches [23]. Within the fine-tuning process, a pre-trained model is used without its top layers, which are replaced by new ones, more appropriate for the new task, and the model is re-trained on the new dataset. In Figure 2 we present the underlying architecture, which has produced state-of-the-art results in various other domains [24, 25, 26]. The red-colored block represents the input layer of the model *(batch-size, image-width, image-height, image-channels)*. The dashed red block represents the image augmentation layers which are considered an essential part of the input layer. The blue-colored block depicts the pre-trained stack of layers of the popular architectures (pre-trained on ImageNet [27]), which are mainly responsible for the feature extraction. The green-colored block depicts the top layers that are selected according to the new task and the two small yellow blocks represent the output of the model.

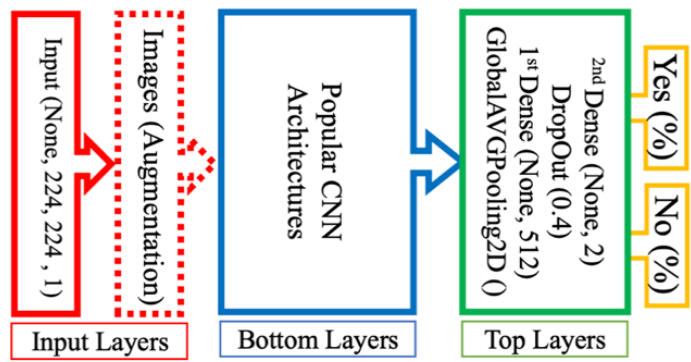

Figure 2. Adopted transfer learning architecture for the classification experiments.

In this context, our experimental methodology relies on a fine-tuning approach exploiting widely used DL models, which were pre-trained on the ImageNet dataset [28]. Within our first-stage experiment we investigated the performance of the following models/architectures: VGG16 [29], DenseNet121 [30], MobileNetV2 [31], ResNet50 [32], InceptionV3 [33], NasNet [34], InceptionResNet [35], as well as the EfficientNet [36].

At this point, it should be highlighted that we use the same hyperparameter configuration for all models, in order to more fairly assess their effectiveness in CTPA-Scan images. Specifically, we have used a $224x224x1$ tensor for each image as input size for training and evaluation of the model. For training, we have chosen Adam as the optimization algorithm [37] with a learning rate of 0.0001. For the top-layers, we used a dense layer with 512 neurons followed by the output layer that holds two outputs. The output layer presents probabilities for two classes, a) "Yes" the image presents PE with $\hat{y}_1$ (%) accuracy and b) "No" the image does not present PE with $\hat{y}_2$ (%) accuracy. We approached the problem as a multiclass classification problem in order to not restrict the prediction of the model to a binary True-False state and to provide more flexibility to the medical experts to reach their decision based on the probability of each output (class). For this reason, we use the SoftMax activation function for the output layer. Moreover, we added a Dropout layer [38] (with a 0.4 ratio) and L2 Regularization [39] (with 0.005) to reduce overfitting of the models and to keep the training progress smooth. Due to the challenging nature of the data (grayscale low-resolution images with high complexity information in their gray shades), a small number of augmentation techniques was exploited, such as: rotation by $10^o$ in random orientation, width and height shift by 5%, zoom by 30%, horizontal flipping and shearing by 20%.

## 2.3 Localization of pulmonary embolism CTPA-Scan images

Localization and OD are some of the most important core applied research tasks in CV and ML science, which focus, not only in the classification of an image to a specific class, but also in the spatial localization of the various entities depicted in the image. Simply put, localization is the approach of finding the position of the classified object in an image and highlighting it with a bounding box, while OD is the process of the detecting and localizing multiple objects (from the same or from multiple different classes) in an image and providing a confidence accuracy for each one. Nowadays, there are many different object detection models which can be categorized into two categories [40]:

- Two-stage detectors: in the first stage, the model proposes important features and in the second stage the model classifies and localizes all important features one by one. The most representative models are R-CNN [41] and its variations/improvements [42, 43].
- One-stage detectors: these propose predicted bounding boxes (objects localization) and their confidence accuracy for each object, directly, without the regional proposal step. The most representative models are YOLO [44] and SSD [45].

Recently, two-stage detectors were considered to provide high localization and object recognition accuracy, whereas the one-stage detectors achieved high inference speed [40]. In 2020 an improvement of the YOLO was proposed, the YoloV4 [46] and its variations [47], which are considered to be the fastest and most accurate object detector on the MS COCO dataset[2] [48]. In addition, they provide variation of the anchor boxes for better detection of small objects, which is very important for the detection of the PE in CTPA-Scan images.

During the training of the model, we exploited some default parameters of the configuration and setup of the YOLOv4. We trained the model with 10,000 iterations, a 0.001 learning rate, a 0.949 momentum and an image size of $416 \times 416 \times 1$ (grayscale). It should be highlighted that we adopted pretrained weights of YOLOv4 to increase the accuracy and the training speed. After many fine-tuning tests, the aforementioned setup produced the best model at a 50% Intersection over Union (IoU)[3].

## 3. Results

This section presents the main outcomes of the two aforementioned experimental stages, which are based on DL approaches. In the first stage, classification of CTPA-Scans images, into two different classes, one for images presenting PE and one for not presenting PE. The validation accuracy curves of each model are depicted in Figure 3, where the predominance of the MobileNet model is clear with the smoothest curve and the highest accuracy, of about 0.91.

---

[2] One of the most popular benchmark datasets in object detection tasks. https://cocodataset.org/

[3] Intersection over Union (IoU) is a metric used in CV science, especially for objects localization in images/videos, where it is calculated the prediction precision by dividing the Intersection area per the Union area of the predicted and base-line bounding boxes. If the IoU of a predicted object is above *X*, then the prediction is considered as True Positive.

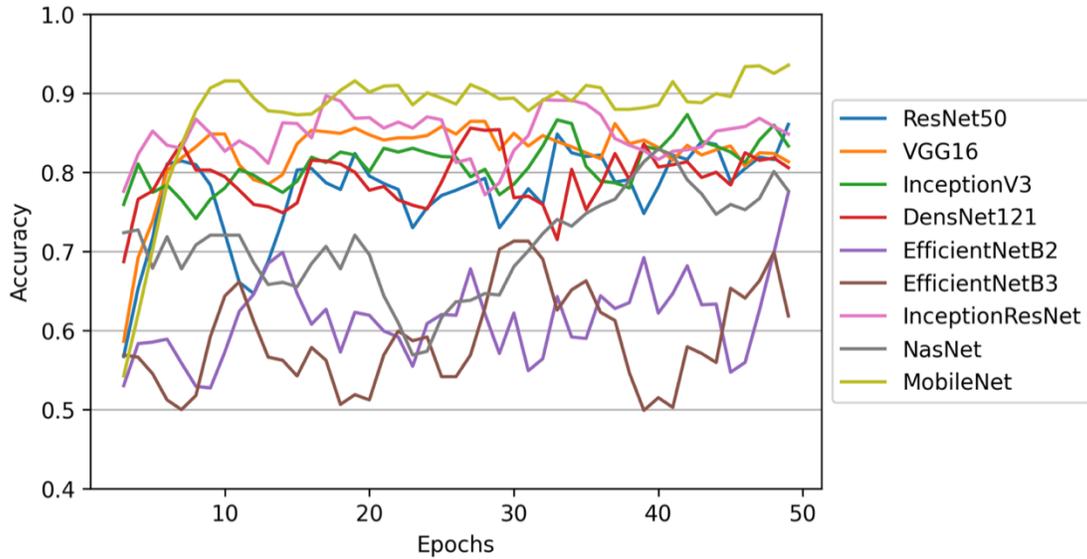

Figure 3. The validation accuracy curves of the assessed models/architectures

Based on the validation accuracy curves (of Figure 3) we ranked all the model in terms of decreasing accuracy. By taking into account and averaging the accuracies of the last 20 epochs of each model, we present their rankings in Table 1. Due to the high fluctuation in the evolution of some curves (e.g. EfficientNet), this approach is considered to produce accurate model ranking. Once again, the predominance of MobileNet is highlighted, followed by the InceptionResNet with a difference of ~7,28%. Although the VGG16 has been ranked in the fifth position, examining the curves in Figure 3 reveals that its evolution is smoother compared to the three models that performed better in terms of accuracy (InceptionResNet, InceptionV3 and ResNet50). At this point it should be highlighted that the best performing model has the fewest learning parameters.

Table 1. The ranking table of the adopted state-of-the-art models/architectures.

| Ranking | Model | Parameters (million) | Validation accuracy (%) |
|---|---|---|---|
| 1 | MobileNet | 3.0 | 91.63 |
| 2 | InceptionResNet | 55.2 | 84.36 |
| 3 | InceptionV3 | 22.9 | 83.52 |
| 4 | ResNet50 | 24.7 | 83.45 |
| 5 | VGG166 | 15.0 | 82.65 |
| 6 | DenseNet | 7.6 | 80.80 |
| 7 | NasNet | 87.0 | 77.35 |
| 8 | EfficientNetB2 | 9.1 | 66.29 |
| 9 | EfficientNetB3 | 12.3 | 61.40 |

In the second experimental stage, localization of the suspect RoI of PE into CTPA-Scan images is examined through an OD approach. We achieved an AV of 68% at 50% IoU, with a quite smooth evolution and a very low Loss, of about 0.065 at the 8,200$^{th}$ epoch. In that epoch, we kept the best model, because the model started overfitting after that point. The training progress of the model is depicted in Figure 4, while in Figure 5 we also present the AP, the average IoU of the predictions and the F1 scores of the model at various IoU thresholds Figure 5. The F1 scores follow the AP scores, with a smooth downhill progress during the increase of the IoU threshold. Also, the AVG IoU curve flattening is clear, per various IoU thresholds for percentages less than 50%, highlighting the need for further training data with more complex patterns. On the other hand, we note an AP score increase of about 9%, in 40% IoU threshold. In many cases this approach could be considered as inappropriate (low overlapping of the predicted area over the ground-truth), however in this scenario it is very important to highlight any suspect RoI, since the ultimate decision will be upon the domain experts (such as Radiologists etc), and the proposed system is developed to assist the domain experts towards the achievement of a more accurate PE diagnosis. To sum-up, the proposed model uses an IoU threshold of 50% as per the literature [49].

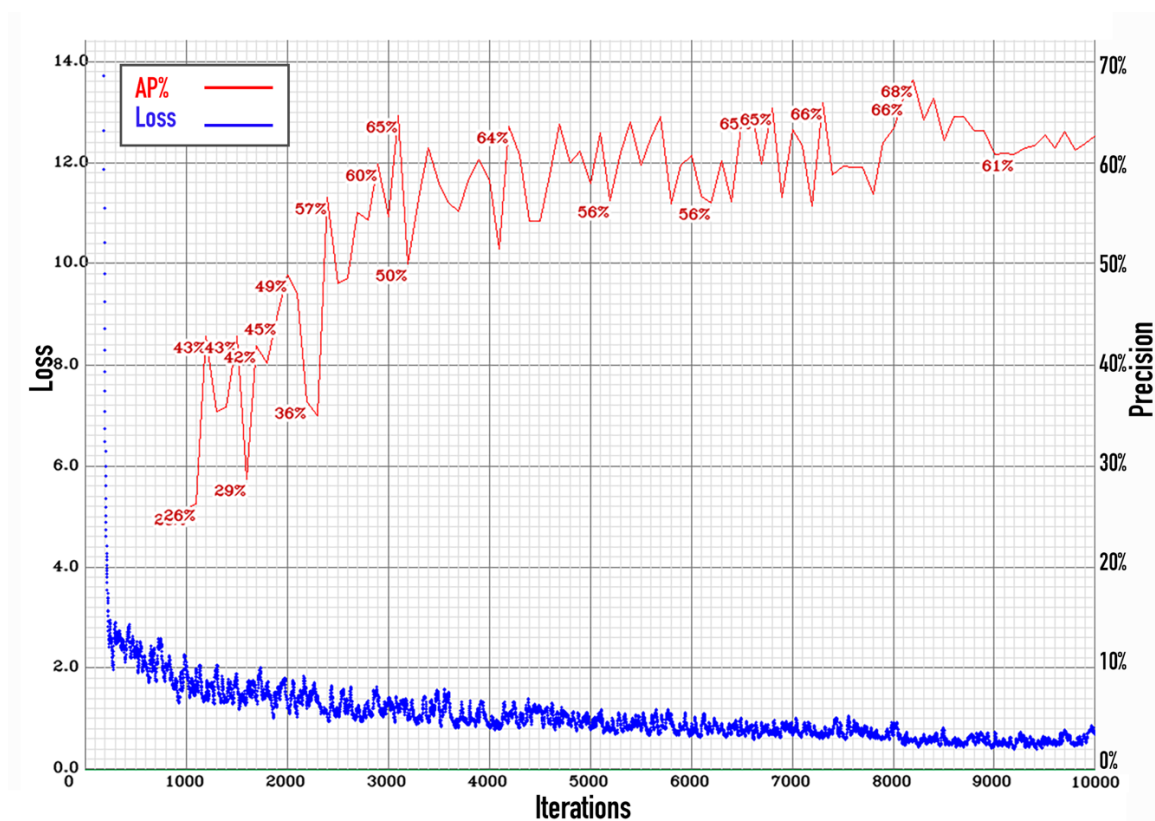

Figure 4. Object detection loss and average precision curves for the localization (object detection) of pulmonary embolism.

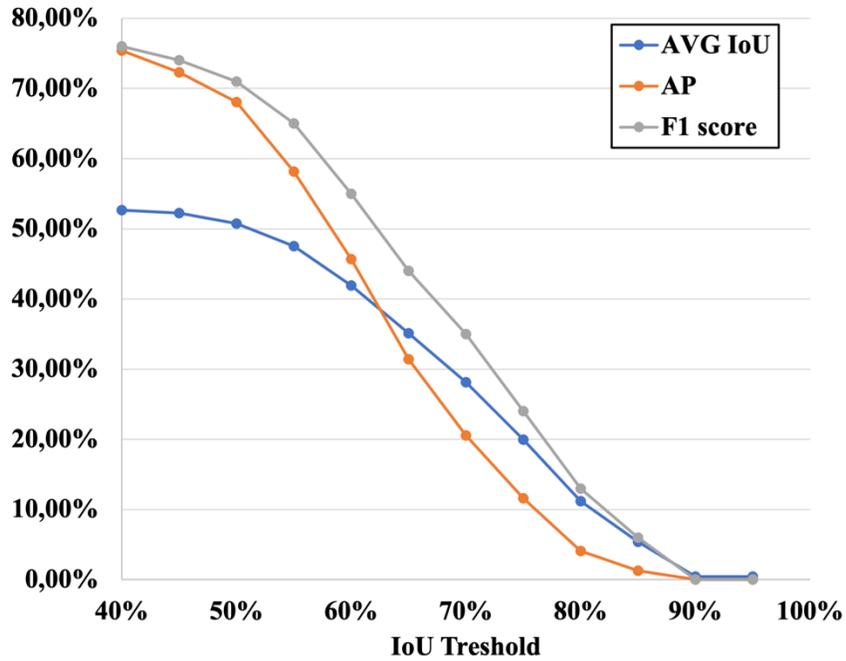

Figure 5. Average precision, F1 score and average IoU of the bounding boxes in different IoU thresholds, in the validation of the DO model.

Figure 6 depicts the results from two different CTPA-Scan image samples, one easy-case for the identification of PE (Figure 6a) and one difficult-case (Figure 6b) for radiologist experts. In Figure 6a, both classification and localization models give very high accuracies. In Figure 6b, the localization of the first RoI of the PE (the smallest one, left-side of the image) is localized with low accuracy (about 0.52) and could generate some confusion in the final decision. However, the existence of PE can be easily confirmed by the combination of the high confidence for the second RoI (0.94 confidence) of the suspect PE and the almost perfect classification accuracy.

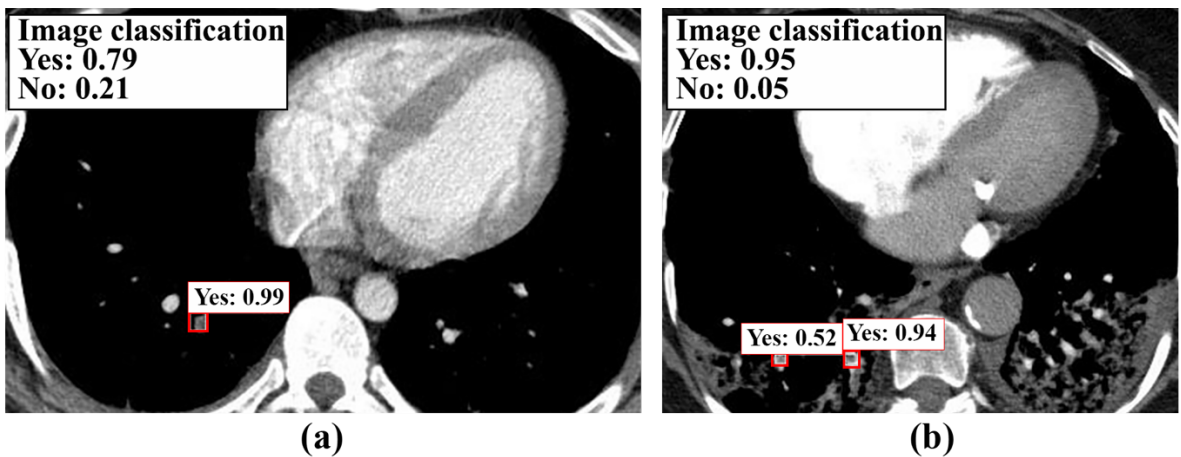

Figure 6. Localization of RoI of PE and identification of PE: a) an easy-case for experts and b) a difficult-case for experts. Figure 6b is the same with Figure 1c.

# 4. Discussion

The short assessment of the classification models highlights that DL approaches with relatively low architecture complexity and low number of learning parameters, such as MobileNet, performed very well in CTPA-Scan images, in contrast to very complex architectures, such as the NasNetLarge or EfficientNet, which achieved lower accuracy in PE identification. This may be due to the fact that this challenge has low number of classes (just two), or due to the nature of the images (low resolution/quality and gray scale). Although this outcome may contradict with the capabilities of those state-of-the-art architectures, which are designed for large scale datasets composed of color images with tens of classes, the adoption of TL methods is considered to be a vital approach for fast-track solutions, instead of developing custom models/architectures, which is a complex and time-consuming process. By observing the outcomes of the OD model, it is clear that the proposed approach provides very good localizations of the PE RoI; this was also confirmed by the Radiology team of the Sismanogleio General Hospital. Even with an AP of 68%, which can be considered as a very accurate approach, the introduced OD model cannot provide perfect overlapping of the predicted area over the ground-truth one. However, as we mentioned earlier, our main goal is to provide a strong suggestion to experts of the area towards utilising these techniques to assist them in making more thoughtful decisions for more accurate identification of PE.

Both classification and OD models provide quite high accuracy and AP respectively. As far as we know (based on the limited existing related literature), this is the first approach of the localization of the RoI of PE in CTPA-Scan images through DL and OD technologies, which achieved 68% AP at 50% IoU. In contrast to related works, we provide to the research community an easy-to-use prototype system (tool)[4] with many future perspectives, which combines both classification and OD models towards more accurate predictions and localization of PE. In addition to that, the system, its training pipelines, the prediction algorithms and the weights of both models (neural networks) are publicly available through a source code management repository, to assist researchers within this domain.

One of the important limitations of the training approach is that we used CTPA-Scans from a single CT-Scan system (software and hardware) and, as we noticed, the quality of CTPA-Scans may vary from system to system. Since, all the included CTPA-Scans had no artifacts or technical deficiencies, our models may present low accuracy in CTPA images that are not of similar quality or produced under the similar protocol. Moreover, as in some samples the two models may slightly disagree over their predictions, the proposed system is aimed at experts of the area (Radiologists etc.) who are expected to use it mainly for experimentation and not for production/fielding purposes.

---

[4] https://tinyurl.com/pulemb-ctscans

In conclusion, we propose a novel prototype tool, which combines DL and CV technologies that achieves high classification accuracy and high AP in the localization of PE RoI in CTPA-Scan images for COVID-19 patients. Moreover, we provide all the sources (models, code, experimental pipelines etc) of this project to actively boost the research of the domain. As a future direction we plan to increase the size of the dataset, test other ML approaches and to provide a more flexible integrated system (interface) to the research community.